\documentclass[preprint,pre,aps,showpacs]{revtex4}
\usepackage{graphicx}
\begin{document}

\title{Model for projectile fragmentation: case study for Ni on Ta, Be and Xe
on Al}

\author{S.Mallik, G.Chaudhuri}
\affiliation{Variable Energy Cyclotron Centre, 1/AF Bidhannagar, Kolkata
700064}
\author{S. Das Gupta}
\affiliation{Physics Department, McGill University,
Montr{\'e}al, Canada H3A 2T8}

\date{\today}

\begin{abstract}

For projectile fragmentation we work out details of a model whose origin
can be traced
back to the Bevalac era.  The model positions itself between the
phenomenological EPAX parametrization and microscopic
transport models like ``Heavy Ion Phase Space Exploration Model''(HIPSE)
and antisymmetrised molecular dynamics(AMD).  We apply the
model to some recent data of projectile fragmentation of Ni on Ta and Be
at beam energy 140 MeV/nucleon and some older data of Xe on Al at
beam energy 790 MeV/nucleon.  Reasonable
values of cross-sections for various composites populated in
the reactions are obtained.

\end{abstract}


\maketitle

\section{Introduction}
In heavy ion collisions, if the beam energy is high enough, the following
scenario can be envisaged.  For a general impact parameter, part of the
projectile will overlap with part of the target.  This is the participant
region where violent collisions occur.  In addition there are two mildly
excited remnants: projectile like fragment (PLF), with rapidity close
to that of the projectile rapidity and target like fragment (TLF) with
rapidity near zero.  The PLF has been studied experimentally, this being
one of the tools for production and identification of exotic nuclei.

The PLF has mild excitation and breaks up into many composites.  Extensive
measurements of cross-sections of composites arising from the break up
of PLF of Ni on Be and Ta were made at Michigan State University\cite{Mocko1}.
Powerful and elaborate calculations for the case of Ni on Be were made
recently using transport model\cite{Mocko2}.  Unfortunately calculations for Ni
on Ta could not be done because this becomes prohibitively large.
One of the main reasons of this venture was to examine if an alternate,
less ambitious but realistic model, could be used to calculate results
for the case of Ni on Ta.  It appears that above a certain beam energy the
model will be in general applicable and is implementable.

Great progress has been made in phenomenological EPAX \cite{Summerer} model which
predicts results for cross-sections.  Our model, we believe is less
phenomenological.  It is grounded in traditional concepts of
heavy ion reaction plus by now, well-known model of multifragmentation.
We describe the basics of the model below.

\section{Model Assumptions}
Imagine that the beam energy is high enough so that using straight line
trajectories one can uniquely define participant,PLF and TLF.  A certain
fraction of the projectile is lost in the participant.  This can be
calculated. What remains of the projectile
is the PLF, moving with velocity close to the beam velocity.
There is a probability of having N neutrons and Z protons in the PLF.  This
probability $P_{N,Z}(b)$ depends upon the impact parameter.  We call
this abrasion.  The abrasion cross-section when there are $N$ neutrons
and $Z$ protons in the PLF is labelled by $\sigma_{a,N,Z}$:
\begin{equation}
\sigma_{a,N,Z}=\int 2\pi bdbP_{N,Z}(b)
\end{equation}
This is stage 1 of the calculation.

An abraded system with $N$ neutrons and $Z$ protons has excitation.  We
characterise this by a temperature $T$ instead.  This will expand and
break up into many excited composites and nucleons.  This  break up
is calculated using a canonical thermodynamic model(CTM) \cite{Das}.
The cross-section
at this stage is called $\sigma_{n,z}^{pr}$.  This is the second stage
of the calculation.  This second stage can be replaced by another
statistical multifragmentation model(SMM) \cite{Bondorf} but the results are expected to
be very similar \cite{Chaudhuri1}.

Lastly we consider composites after stage 2.  These have a
temperature and can evaporate light particles like neutrons, protons,
alphas etc.  This can deplete a nucleus with neutron and proton numbers
$n$ and $z$ that was obtained after stage 2
but there is a compensation also by feeding from higher
mass nuclei.

\section{Calculational Details}
Consider the abraison stage.  The projectile hits the target.  Use
straightline geometry.  We can then calculate the volume of the
projectile that goes into the participant region (eqs. A.4.4
and A.4.5 of \cite{Dasgupta1}).  What remains
in the PLF is $V$.  This is a function of $b$, the impact parameter.
If the original volume of the projectile
is $V_0$, the original number of neutrons is $N_0$ and the original
number of protons is $Z_0$ then the average number of neutrons in
the PLF is $<N(b)>=(V(b)/V_0)N_0$ and the average number of protons is
$<Z(b)>=(V(b)/V_0)Z_0$.  These will usually be non-integers.
Since in any event
only integral numbers for neutrons and protons can materialise in the PLF,
we have to guess what is the distribution of $N,Z$ which produces these
average values.

Two distributions immediately come to mind.  One is a minimal distribution
model.  Let $<N(b)>=N_{min}(b)+\alpha$ where $\alpha$ is less than 1.  We can
also define $N_{max}(b)=N_{min}(b)+1$.  We assume that $P_N(b)$ is zero
unless
$N(b)$ is $N_{min}(b)$ or $N_{max}(b)$.  The distribution is narrow.  We then
get $P(N_{max}(b))=\alpha$ and $P(N_{min}(b))=1-\alpha$.
From $<Z>$ we can similarly
define $P_Z(b)$.  Together now we write $P_{N,Z}(b)=P_{N}(b)P_Z(b)$.
This is the
$P_{N,Z}(b)$ of the previous section(eq.(1)).

The alternative is a binomial distribution which has a long tail.  Now
$P_N(b)$ is defined by $P_N(b)=(^{N_0}_N)(occ(b))^{N}(1-occ(b))^{N_0-N}$
(see also \cite{Gaimard}). Here
$occ(b)=V(b)/V_0$. Similarly we can define $P_Z(b)$ for binomial distribution. We can take $P_{N,Z}(b)=P_N(b)P_Z(b)$.
The binomial distribution would be appropriate if
the projectile is viewed as a collection of non-interacting neutron and
proton gas with constant density throughout its volume.  This is
oversimplification and we find the binomial distribution is too long tailed.
For very peripheral collision (with only 1 or 2 nucleons
lost to the participant)
the temperature of the PLF should be nearly zero and and $\sigma_{a,N,Z}$
can be directly confronted with data.
The calculation gives a far too wide distribution.  The same test applied
to the minimal distribution model shows that it errs on being too narrow.
Here we show results calculated with minimal distribution which is easier
to work with.

The limits of integration in eq.(1) are $b_{min}$ and $b_{max}=R_{target}+
R_{projectile}$. For $b_{min}$ we have either 0 (if the projectile is
larger than the target) or $R_{target}-R_{projectile}$ (if the target
is larger than the projectile, in this case at
lower value of $b$ there is no PLF left).  In evaluating eq.(1) we
replace integration by a sum.  The cross-sectional area between $b_{min}$
and $b_{max}$ is divided into $M$ rings of equal cross-sections
and $P_{N,Z}(b)$ is evaluated at midpoints between radii
of successive rings.
For Ni on Be we use $M$=20, for Ni on Ta we use $M$=100 and for
Xe on Al we use $M$=200.

Now we come to the second stage of the calculation.
The abraded system of $N,Z$ nucleons will have an excitation which we
characterize by a temperature $T$.  Previous experiences with projectile
fragmentation lead us to expect a temperature around 5 MeV.  In this work
we fix the temperature from a fit to the data.  This will be explained
soon.
The excitation and hence the temperature of the abraded system owes its
origin to several factors:
deviation from spherical shape when abrasion happens: migration of nucleons
from the participant zone etc..  Estimating the temperature from a
more basic calculation is beyond the scope of this model.

The abraded system with $N,Z$ and a temperature $T$ will break up
into many composites and nucleons.  We use the canonical
thermodynamic model (CTM) to calculate this break up.  As this has
been described many times \cite{Das,Chaudhuri2} we merely specify the composites it
can break into.  It can break up into neutrons, protons,$^2$H ground
state,$^3$H ground state, $^3$He ground state, $^4$He ground state
and heavier nuclei in ground and excited states.  For these heavier
nuclei the following approach is taken. We use the liquid drop mass
formula which defines neutron and proton drip lines.  All nuclei
within drip lines are included.  Excited states of these nuclei are
included using a density of states derived from a Fermi-gas model.
The hot abraded system expands.  The dissociation into various
nuclei according to thermodynamics is calculated at this larger
volume.  Although a range of freeze-out volumes were considered we
show only results for freeze-out volume $3V_0$ where $V_0$ is the
normal nuclear volume.  A larger volume is normally used for break
up of the participant zone but for disintegration of PLF the value
3$V_0$ was used in the past with good success \cite{Chaudhuri4,Chaudhuri5}.

If we have, after abrasion, a system $N,Z$ at temperature $T$, CTM allows
us to compute the average population of the composite with neutron number
$n$ and proton number $z$ when this system breaks up.
Denote this by $n_{n,z}^{N,Z}$.
It then follows summing over all the abraded $N,Z$ that can
yield $n,z$ the primary cross-section for $n,z$ is
\begin{equation}
\sigma_{n,z}^{pr}=\sum_{N,Z}n_{n.z}^{N,Z}\sigma_{a,N,Z}
\end{equation}
This finishes stage 2 of the calculation.

The composite $n,z$ obtained after CTM is at temperature $T$.  It
can $\gamma$-decay to shed its energy but may also decay by light
particle emission to lower mass nuclei.  On the other hand some
higher mass nuclei can decay to this composite.  We include
emissions of $n,p,d,t,^3$He and $^4$He. Particle decay widths are
obtained using the Weisskopf's evaporation theory \cite{Weisskopf}. Fission is
also included as a de-excitation channel though for the nuclei of
mass $<$ 100 its role will be quite insignificant.

Once the emission widths ($\Gamma$'s) are known, it is required to
establish the emission algorithm which decides whether a particle is
being emitted from the compound nucleus. This is done \cite{Chaudhuri3} by first
calculating the ratio $x=\tau / \tau_{tot}$ where $\tau_{tot}= \hbar
/ \Gamma_{tot}$, $\Gamma_{tot}=\sum_{\nu}\Gamma_{\nu}$ and $\nu =
n,p,d,t,He^3,\alpha,\gamma$ or fission and then performing
Monte-Carlo sampling from a uniformly distributed set of random
numbers. In the case that a particle is emitted, the type of the
emitted particle is next decided by a Monte Carlo selection with the
weights $\Gamma_{\nu}/\Gamma_{tot}$ (partial widths). The energy of
the emitted particle is then obtained by another Monte Carlo
sampling of its energy spectrum. The energy, mass and charge of the
nucleus is adjusted after each emission and the entire procedure is
repeated until the resulting products are unable to undergo further
decay. This procedure is followed for each of the primary fragment
produced at a fixed temperature and then repeated over a large
ensemble and the observables are calculated from the ensemble
averages. The number and type of particles emitted and the final
decay product in each event is registered and are taken into account
properly keeping in mind the overall charge and baryon number
conservation. This is the third and last stage of the calculation.
The details of how we do this are given in \cite{Chaudhuri2}.

\section{Some General Features}

There is one parameter in the model: the temperature $T$.  As
already mentioned: there are at least two reasons why the PLF has an
excitation. The abraded remnant did not start with a spherical shape
and one expects some migration from the participant.  Without a
calculation at a more fundamental level it is not possible to
calculate the excitation.  We do not deal with excitation energy as
such and characterise the system by a temperature $T$.  It is
expected that the temperature should be fairly constant as a
function of the impact parameter $b$ (see also \cite{Mocko2}) except for very
peripheral collisions where it will rapidly drop to zero. To keep
the model as parameter free as possible we use one temperature for
all $b$.  There is a price to pay.  For very peripheral collisions
(loss of only one or two nucleons to participants) we can not expect
reasonable results.  We will demonstrate this later.

The projectile-target combinations we have chosen highlight different aspects.
Consider Ni on Be.  The projectile is significantly larger than the target.
In such a case, the abraded projectile has a lower limit on $N,Z$ (as Be
can drive out only some nucleons, not all).  For $^{64}$Ni on Be
the abraded projectile fragment has, on the average 22 neutrons
and 17 protons for $b\approx 0$
(for larger impact parameter $b$ it can have more neutrons
and protons but not less).  But significant cross-sections
exist for composites with z=8,9,10 etc.  These therefore must arise
from canonical model break-up (stage 2) of an abraded system (stage 1).
On the other hand for Ni on Ta (projectile smaller than target) the abraded
system itself covers most of the range of composites seen in the experiment.
The role of the second stage (eq.(2)) is to modify the cross-sections.
The case of $^{127}$Xe on Al highlights another aspect.  Here
the abraded systems are very neutron rich and must shed many neutrons
(stage 2) before comparison with experiment can be done.

For the case of Xe on Al at 790 MeV/nucleon obvious arguments can be
given for defining participants and
spectators using straightline geometry.  At 140 MeV/nucleon
(Ni on Be and Ta) we are probably near the lower limit where this is still
an acceptable approximation.  An interesting question is: do we expect
the same temperature.
We fix the temperature from a fit to the experimental cross-sections.
As there are many many cross-sections, for fixing the temperature we examine
calculated and experimental values of summed cross-sections:
$\sigma_z\equiv\sum_n\sigma(n,z)$
and $\sigma_a\equiv\sum_{n+z=a}\sigma(n,z)$.  We find that both for Ni
on Be and Ta at 140 MeV/nucleon and for Xe on Al at 790 MeV/nucleon
we are led to a value of $T\approx 4.25$ MeV.  Results are given in the
following sections.

At Bevalac where experiments were at higher energies, straight line
geometry was used to define participants and spectators down to 250
MeV/ nucleon, the lowest energy for which data are available \cite{Gosset}.

\section{Temperature Extraction}

We compute total charge cross-sections $\sigma_z=\sum_n\sigma(n,z)$
and total mass cross-sections $\sigma_a=\sum_{n+z=a}\sigma(n,z)$
and compare
with data for $^{58}$Ni on $^9$Be(fig.1) and $^{181}$Ta(fig.2), $^{64}$Ni on
$^9$Be(fig.3) and $^{181}$Ta(fig.4) and $^{127}$Xe on
$^{27}$Al(fig.5). We show results for $T$=3.25 MeV, 4.25 MeV and 5.25 MeV.
The intermediate value of 4.25 MeV fits the multitude of data better
than the other two.  For brevity we do not show results with other
values of temperature in this range.
Recall that beam energy has changed from 140
MeV/nucleon to 790 MeV/nucleon and a variety of target-projectile
combination has been used  but the temperature in the PLF has not moved much
which is in accordance with the model of limiting fragmentation.

The results in figs.1-5 do not include very peripheral collisions.
For very peripheral collisions lower temperatures should be more
appropriate.  The use of one temperature for all impact parameters
renders our calculation
for very peripheral collisions quite inaccurate.  We show this
in fig.6 where for $^{58}$Ni+$^9$Be we use the same temperature 4.25 MeV
for all impact parameters.  Beyond $z$=25 our calculation underestimates
cross-sections.  With $T$=4.25 MeV nuclei produced very close to$^{58}$Ni
by abrasion
are losing too many nucleons by secondary decay.  At a lower $T$ this
would get cut down.  In this work, from now on, all the results we show
pertain to nuclei with at least two nucleons removed from the projectile.
In later work we hope to improve upon this.  This most likely will
require not only a profile in temperature but also a more sophisticated
model for abrasion.

\section{More Results}
We continue to show results of calculation and compare with experimental data.
All calculations are done with $T$=4.25 MeV and freeze-out volume $V=3V_0$.
The examples shown were picked at random.  We pick an isotope characterised
by a value of $z$ and plot cross-sections for this $z$ for different
values of $n-z$. Fig.7 shows results for the case of $^{58}$Ni on
$^9$Be, fig.8 for $^{58}$Ni on $^{181}$Ta, fig.9 for  $^{64}$Ni on
$^9$Be, fig.10 for $^{64}$Ni on $^{181}$Ta and fig.11 for $^{129}$Xe on
$^{27}$Al.  There are no adjustable parameters any more and the
calculated values of the cross-sections are pleasingly close to experimental
values.  For a given $z$ the general shapes of cross-sections as a function
of ($n-z$) are reproduced but in some cases better mapping would be
desirable.

The topic of isoscaling has been much discussed in recent times.
We examine if isoscaling follows from our calculation.
We know of no obvious reasons why this feature should emerge from this
model but it does(fig.12).   Let $\sigma_2(n,z)$ be the cross-section
for producing the nucleus $n,z$ in the reaction  $^{64}$Ni+$^9$Be and
$\sigma_1(n,z)$  be the cross-section
for producing the same nucleus in the reaction  $^{58}$Ni+$^9$Be.
Let $R_{21}(n,z)=\sigma_2(n,z)/\sigma_1(n,z)$.  Experimentally
log of $R_{21}(n,z)$ falls on a straightline as a function of $n$ for
fixed $z$ and on a different straightline as a function of $z$ for
fixed $n$.  This is called isoscaling.  Fig.12 shows that isoscaling
emerges from this model but the slopes of log of $R_{21}$ are overestimated.

If one is looking at isoscaling only and has many more adjustable
parameters, better fits to isoscaling data are possible \cite{Chaudhuri2}.  Our
objective here is to look at many other data also simultaneously and
we do not have any flexibility. In a recent paper, for the case of
$^{58}$Ni and $^{64}$Ni on $^9$Be isoscaling parameters were
calculated using the HIPSE model \cite{Yao}.

As our last example we consider the production of Si isotopes from
the reaction $^{48}$Ca on $^9$Be at beam energy 140 MeV/nucleon.
This was looked at before \cite{Chaudhuri6,Tsang}.  There the relative values of
cross-sections were calculated using a canonical or a grand
canonical model where the temperature was adjusted to get the best
fit.  For absolute values another constant was needed which was
adjusted.  Here we show (fig.13) absolute values of cross-sections
of Si isotopes with $T$=4.25 MeV and $V=3V_0$ as in all our reported
calculation above.  In expt. the maximum yield is at $n$=16, we get
it at $n$=17. The absolute values of the cross-sections at higher
yield points agree very well but the shape of the theoretical curve
is steeper where the cross-sections are very small.

Several modifications to the model of PLF fragmentation developed here
can be considered.  One would be a
more rigorous choice of $P_{N,Z}(b)$ (eq.(1)).  Another would be variation
of the temperature $T$ in very peripheral collisions.

\section{Summary and Discussion}

Calculations reported here suggest that for beam energy upwards of 140
MeV/nucleon an implementable model for projectile fragmentation gives
reasonable results for cross-sections of end products.
One needs to do an impact parameter integration; at each impact
parameter an abraded nucleus is formed at a temperature of about
4.25 MeV.  This expands to about three times its original volume
and then breaks up thermodynamically into smaller but still hot nuclei.
These can further boil off very light particles reaching the end stage.
It is rather quick to calculate the abrasion and the thermodynamic
break up.  Calculating the evaporation of light particles at the last
stage takes more time.  However we have found that since in the last stage
usually there is both a loss and a gain in the population of many composites
even without the last stage one has an acceptable measure of cross-sections.

While we have reasonable agreements with many data considered here it is
desirable to push the model for improvements.  Two obvious goals will
be: to find a more sophisticated model of abrasion specially at the low
energy end and to build, on physics ground, dependence of temperature on
impact parameter for very peripheral collisions.  We plan to work on these.

\section{Acknowledgements}
This work was supported in part by Natural Sciences and Engineering
Research Council of Canada.

\begin{figure}
\includegraphics[width=6.0in,height=4.5in,clip]{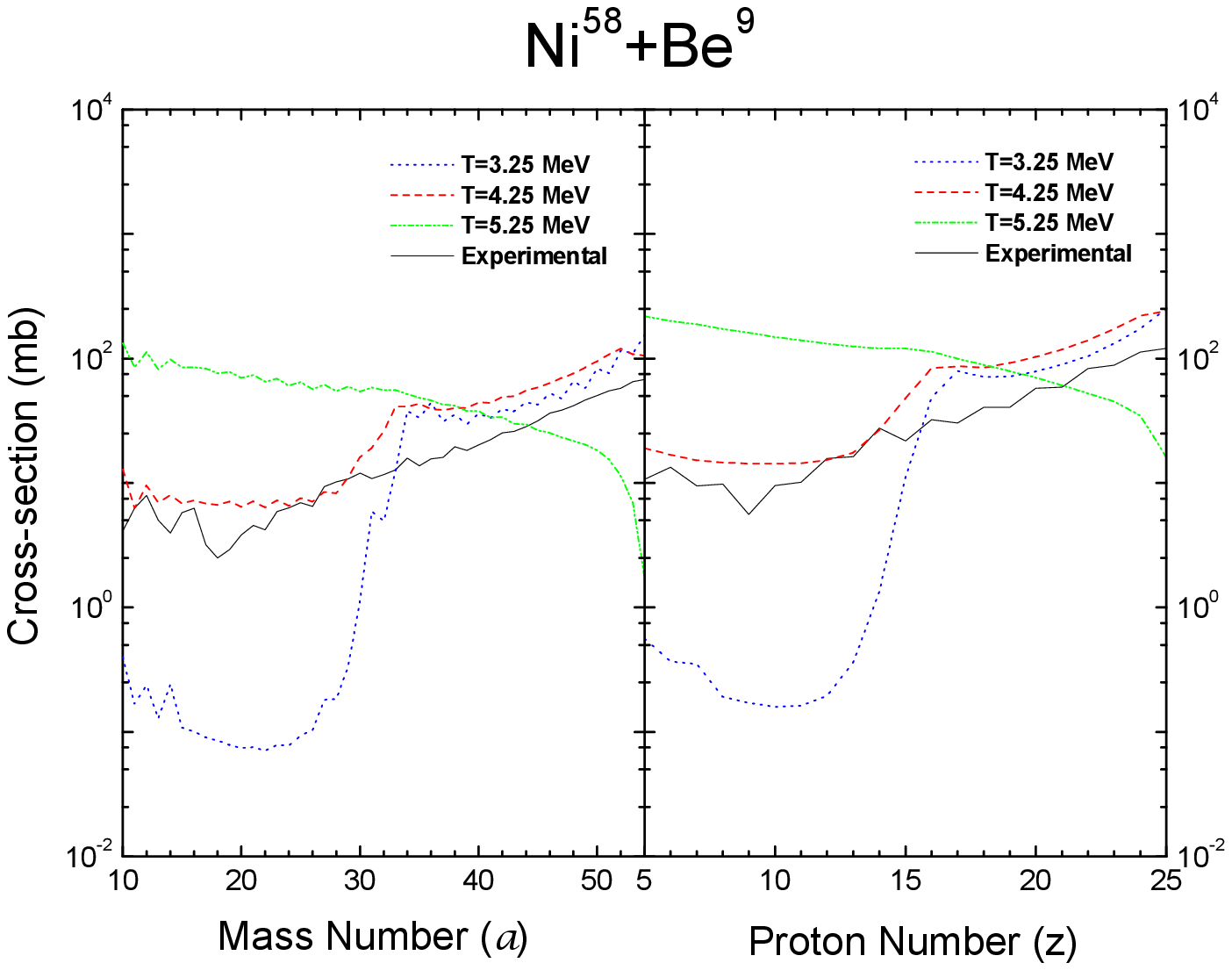}
\caption{ Total mass(left panel) and total charge(right panel)
cross-section distribution for the $^{58}$Ni on $^{9}$Be reaction.
The left panel shows the cross-sections as a function of the mass
number, while the right panel displays the cross-sections as a
function of the proton number. The theoretical results at T=3.25 MeV
(dotted line), 4.25 MeV (dashed line) and 5.25 MeV (dash dotted
line) are compared with the experimental data (solid line).}
\label{fig1}
\end{figure}

\begin{figure}
\includegraphics[width=6.0in,height=4.5in,clip]{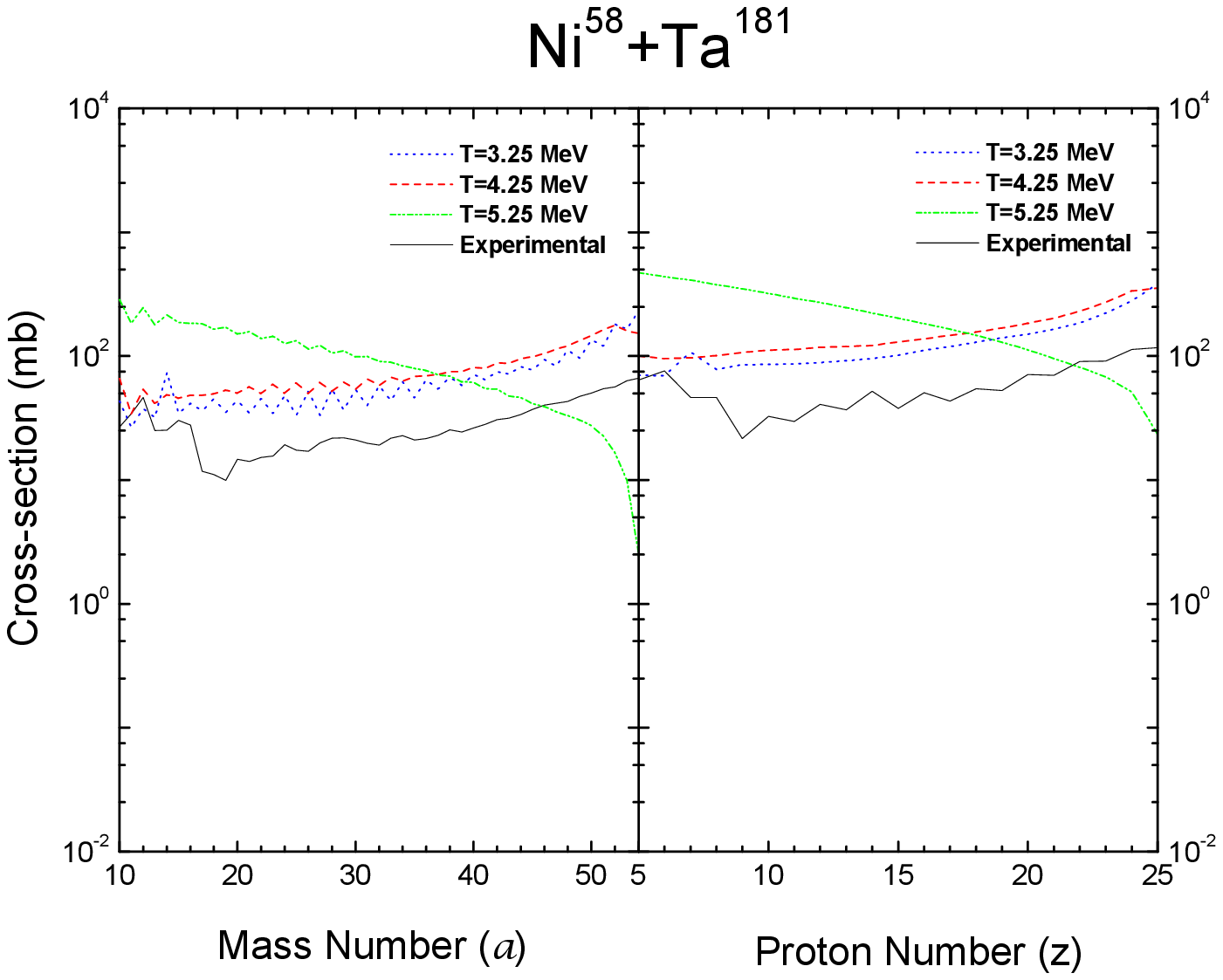}
\caption{ Same as Fig. 1 except that here the target is $^{181}$Ta  instead of $^{9}$Be.} \label{fig2}
\end{figure}

\begin{figure}
\includegraphics[width=6.0in,height=4.5in,clip]{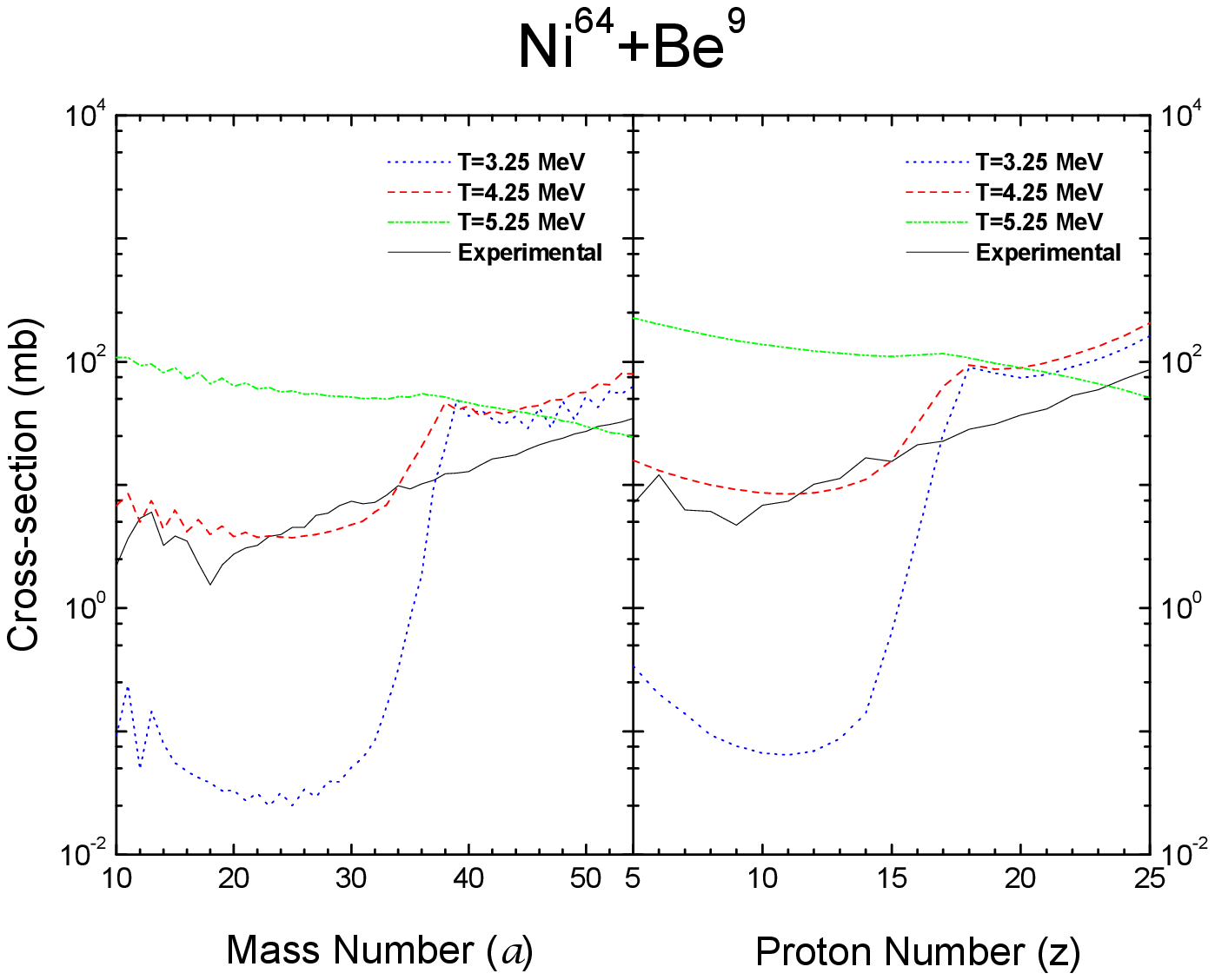}
\caption{ Same as Fig. 1 except that here the projectile is $^{64}$Ni  instead of $^{58}$Ni.} \label{fig3}
\end{figure}

\begin{figure}
\includegraphics[width=6.0in,height=4.5in,clip]{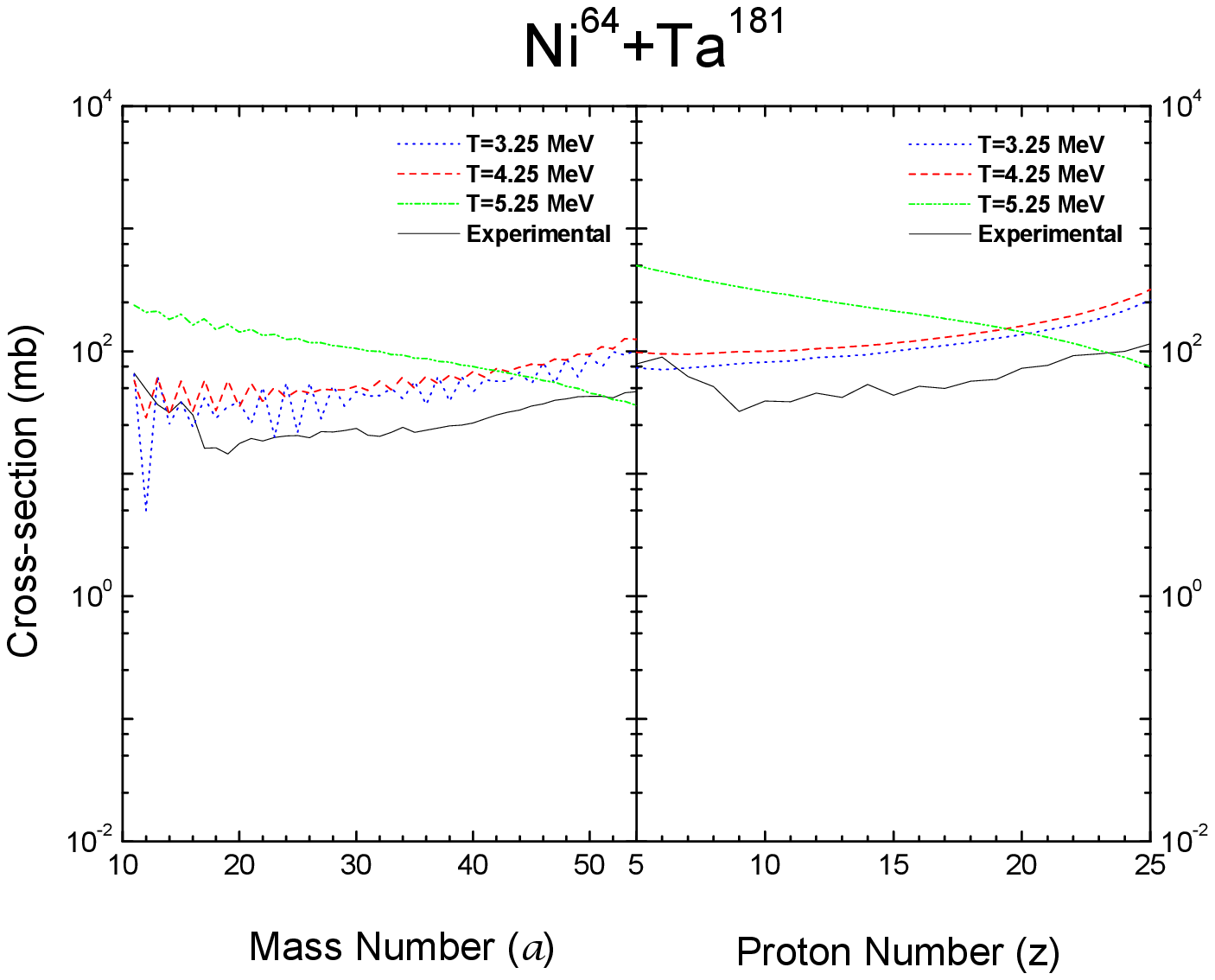}
\caption{ Same as Fig. 3 except that here the target is $^{181}$Ta  instead of $^{9}$Be.} \label{fig4}
\end{figure}

\begin{figure}
\includegraphics[width=3.0in,height=4.0in,clip]{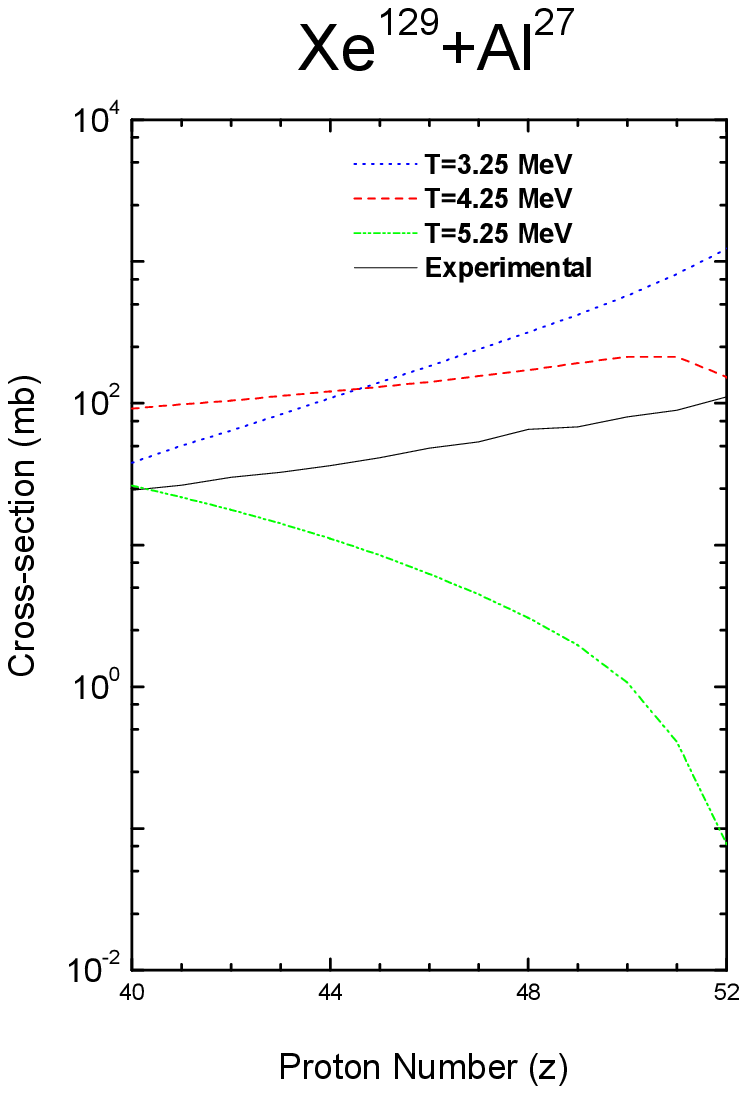}
\caption{ Total charge cross-section distribution for
the $^{129}$Xe on $^{27}$Al reaction. The theoretical results at
T=3.25 MeV (dotted line), 4.25 MeV (dashed line) and 5.25 MeV (dash dotted line) are compared with the experimental data (solid line).}
\label{fig5}
\end{figure}

\begin{figure}
\includegraphics[width=6.0in,height=4.5in,clip]{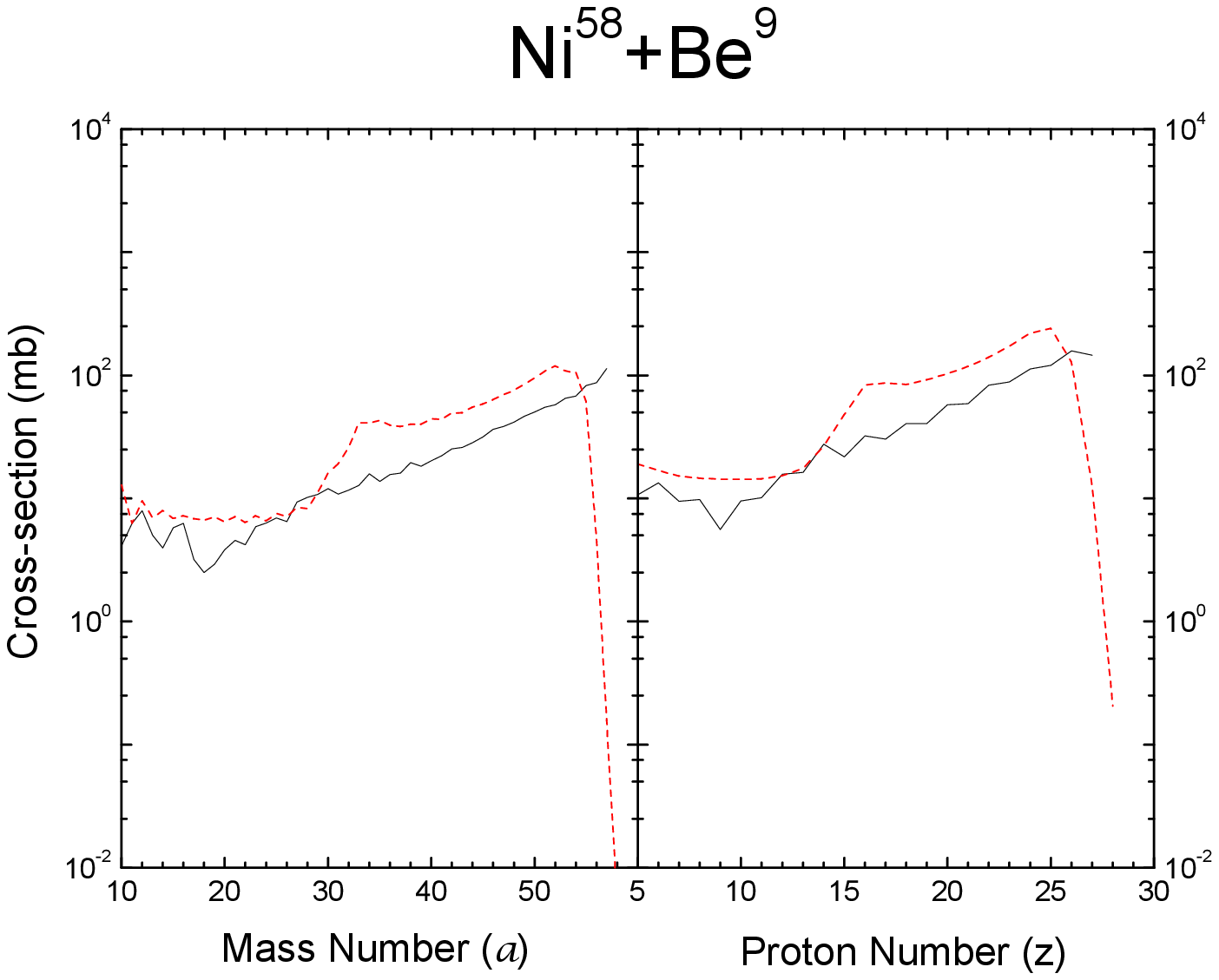}
\caption{Total mass(left panel) and total charge(right panel) cross-section distribution for the $^{58}$Ni
on $^{9}$Be reaction including the regions coming from the very
peripheral collisions. The left panel shows the cross-sections as
function of mass number up-to A=58 (i.e. projectile mass) , while
the right panel displays cross-sections as function of proton number
up-to Z=28 (i.e. proton number of projectile) . The theoretical
result at T=4.25 MeV (dashed line) is compared with the experimental
data (solid line). As, stated in the text, very peripheral collision should have much lower temperature. The discrepancy between theory and experiment near the end is due to the fact that the same T=4.25 MeV is used even for very peripheral collisions. The evaporative loss from the primary is far too great.  } \label{fig6}
\end{figure}

\begin{figure}
\includegraphics[width=6.0in,height=4.5in,clip]{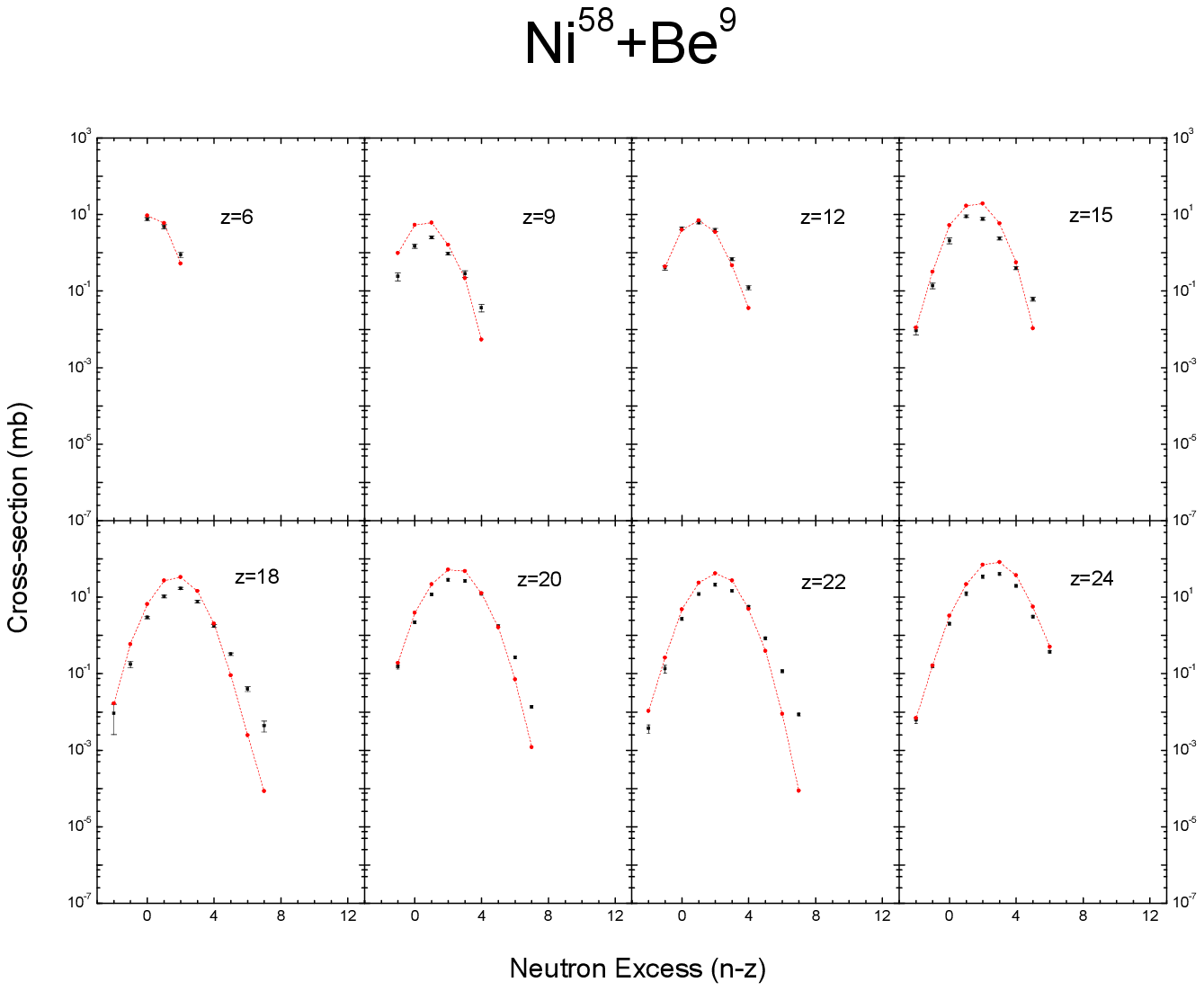}
\caption{Theoretical isotopic cross-section distribution (circles
joined by dashed lines) for $^{58}$Ni on $^{9}$Be reaction compared
with experimental data (squares with error bars).The temperature
used for this calculation is 4.25 MeV.} \label{fig7}
\end{figure}

\begin{figure}
\includegraphics[width=6.0in,height=4.5in,clip]{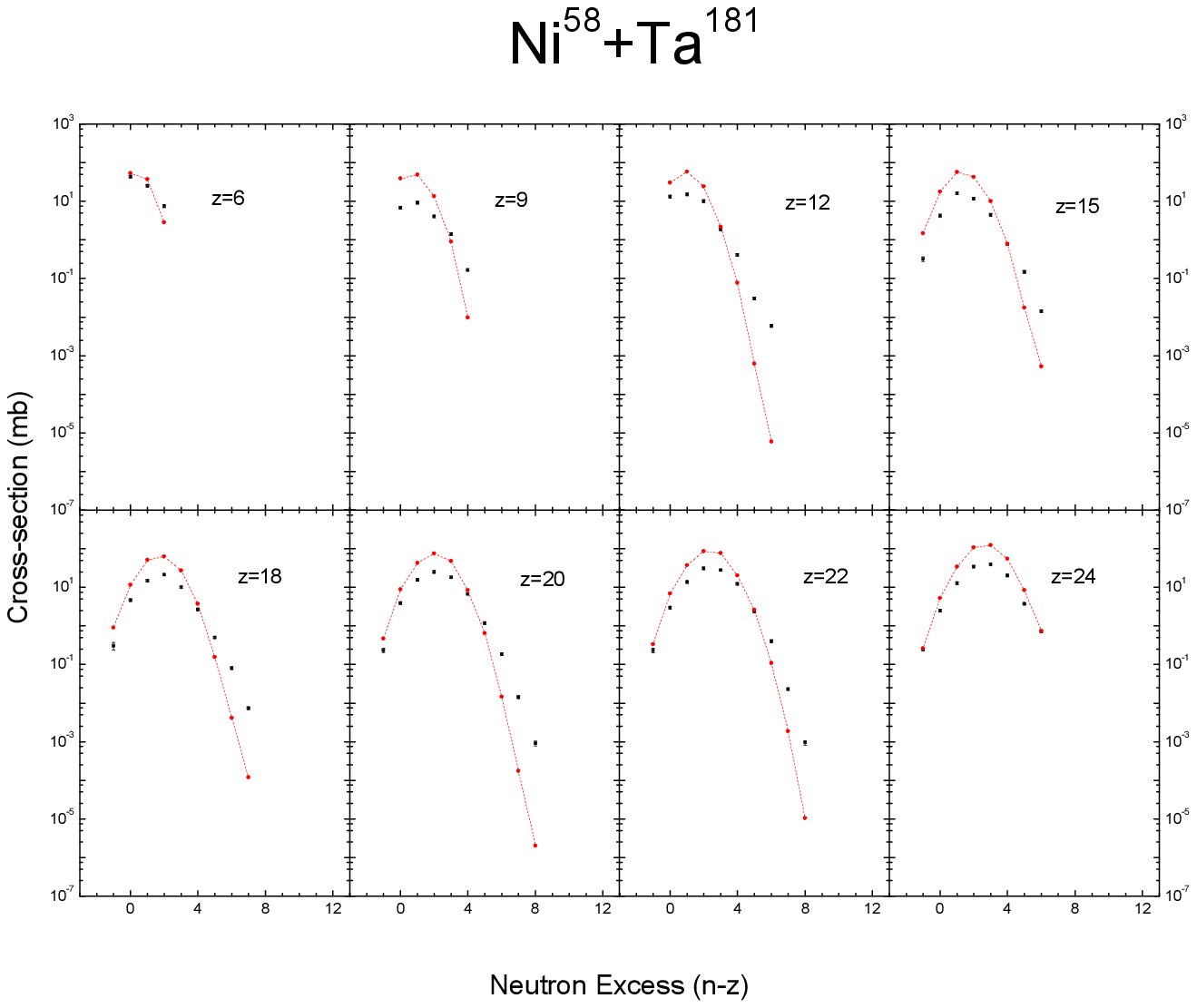}
\caption{ Same as Fig. 7 except that here the target is $^{181}$Ta  instead of $^{9}$Be. The theoretical calculation is done at a temperature T=4.25 MeV.} \label{fig8}
\end{figure}

\begin{figure}
\includegraphics[width=6.0in,height=4.5in,clip]{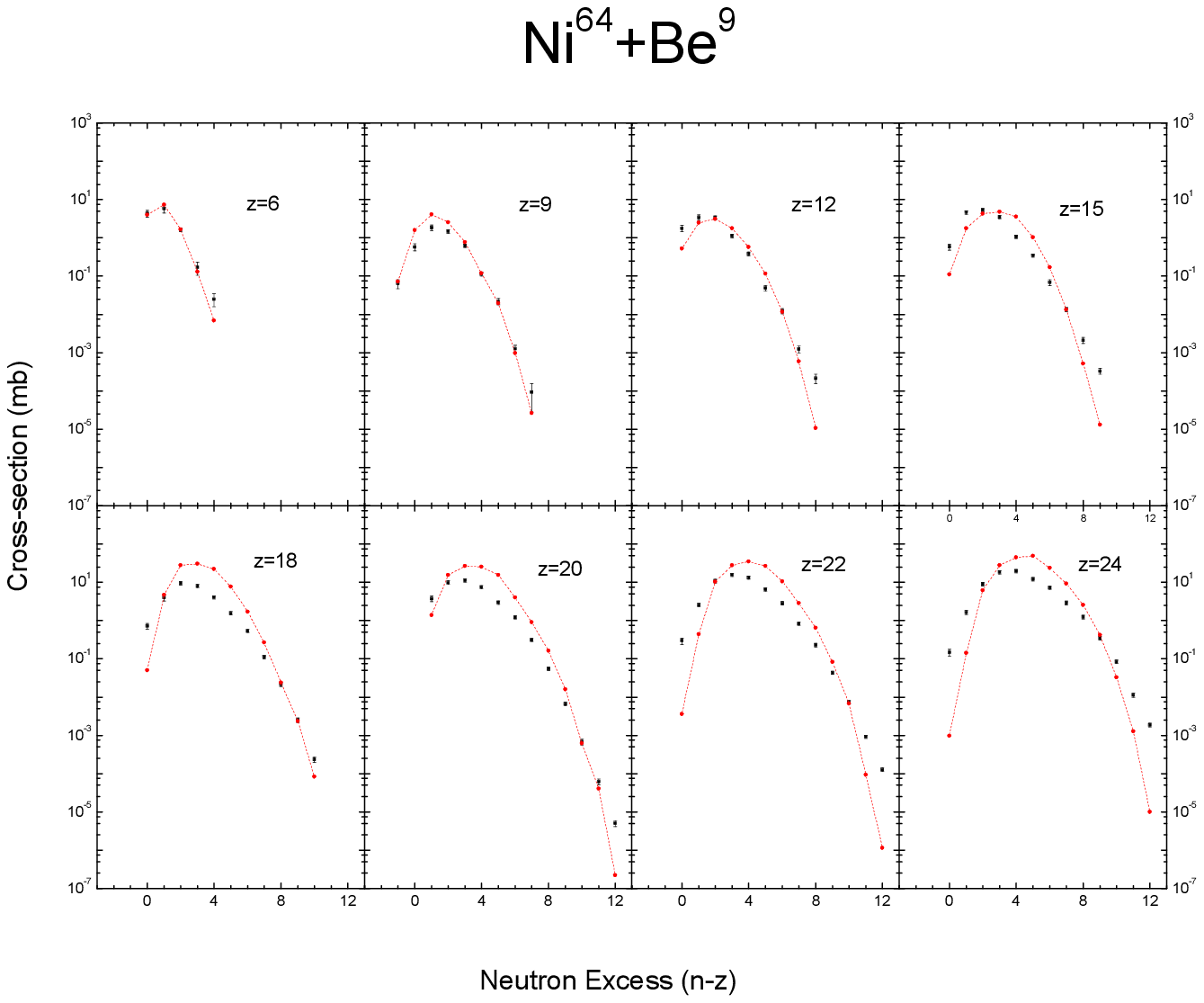}
\caption{  Same as Fig. 7 except that here the projectile is $^{64}$Ni  instead of $^{58}$Ni. The temperature used for this calculation is 4.25 MeV.} \label{fig9}
\end{figure}

\begin{figure}
\includegraphics[width=6.0in,height=4.5in,clip]{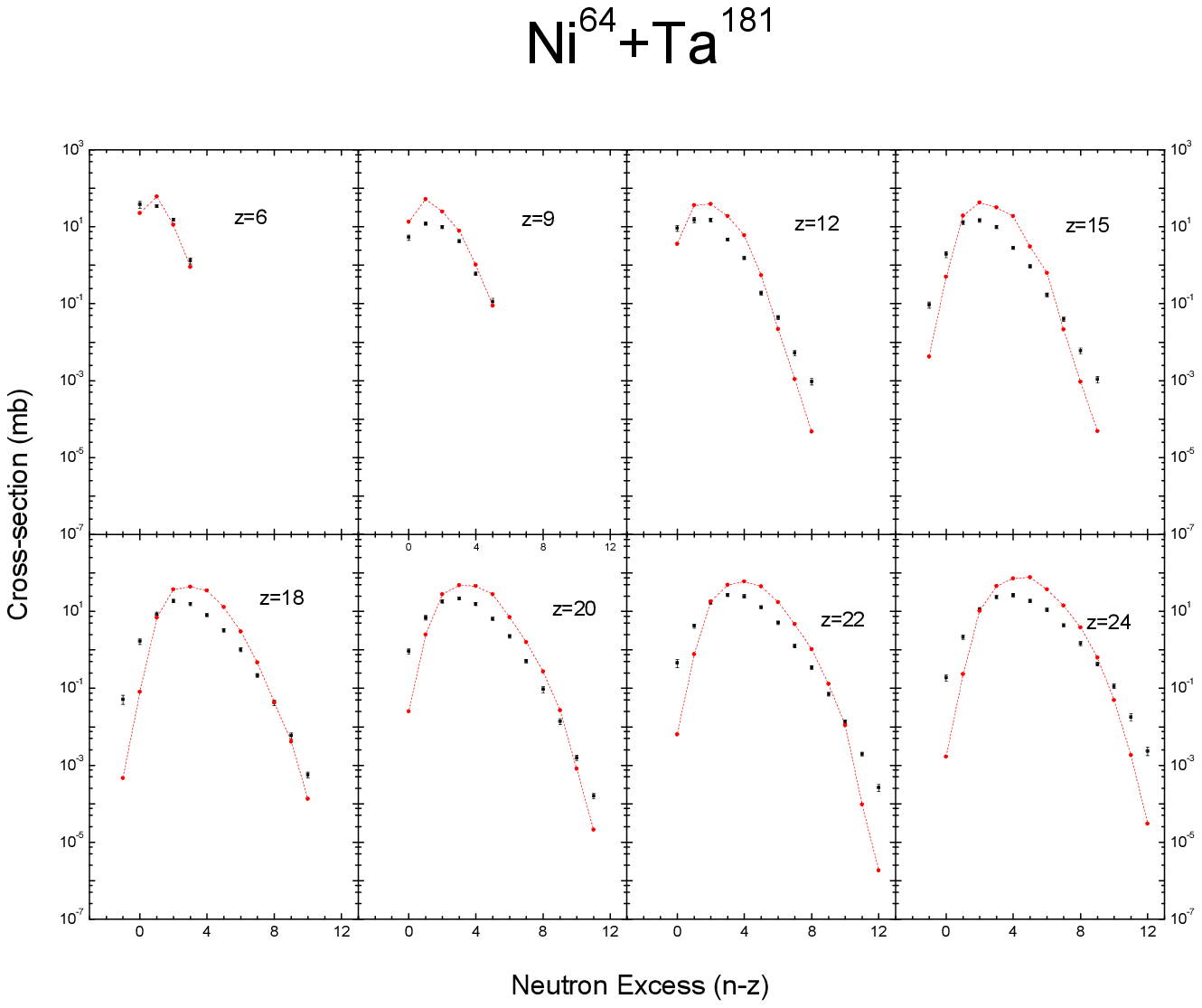}
\caption{ Same as Fig. 9 except that here the target is $^{181}$Ta  instead of $^{9}$Be. The theoretical calculation is done at a temperature T=4.25 MeV.} \label{fig10}
\end{figure}

\begin{figure}
\includegraphics[width=3.0in,height=4.0in,clip]{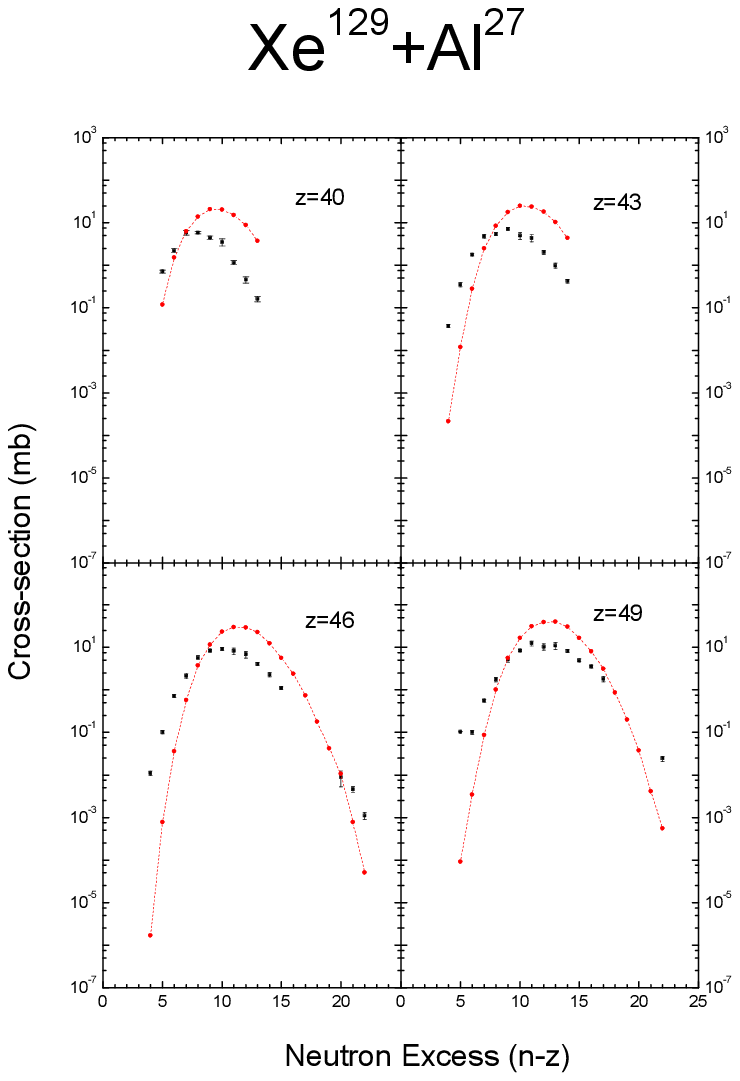}
\caption{ Same as Fig. 7 except that here the reaction is $^{129}$Xe
on $^{27}$Al  instead of $^{58}$Ni on $^{9}$Be. The temperature used
for this calculation is 4.25 MeV.} \label{fig11}
\end{figure}

\begin{figure}
\includegraphics[width=3.0in,height=4.0in,clip]{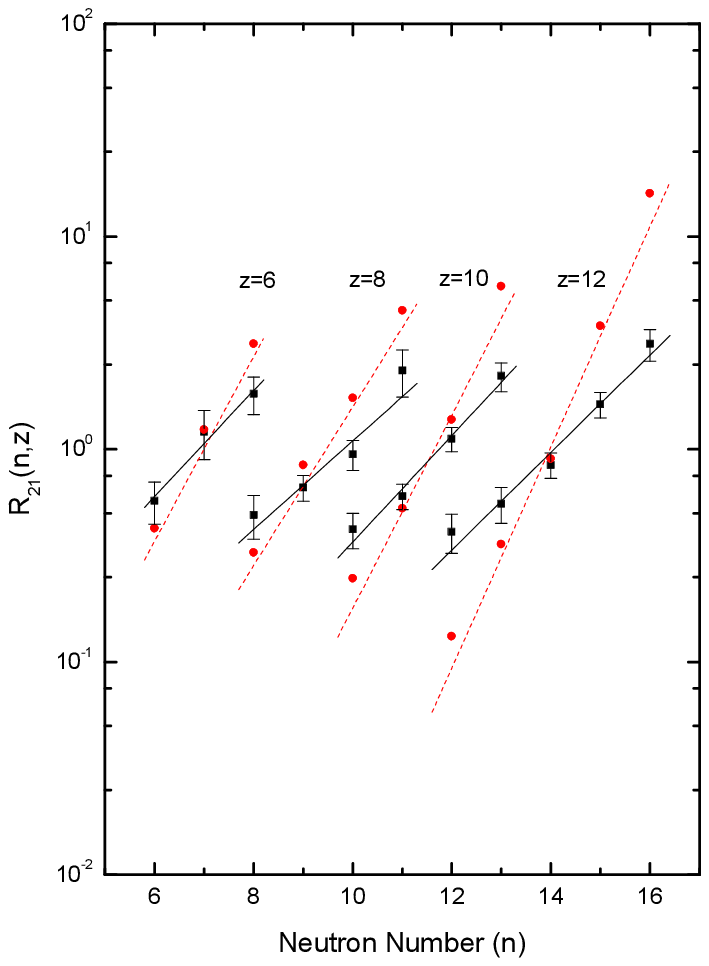}
\caption{ Theoretical ratios of cross-section (circles) of producing
the nucleus(n,z) where reaction 1 is $^{58}$Ni on $^{9}$Be and
reaction 2 is $^{64}$Ni on $^{9}$Be compared with the ratios of the
experimental cross-sections of the same two reactions. The dashed
and solid lines are the best linear fits of the theoretical and
experimental ratios respectively.} \label{fig12}
\end{figure}

\begin{figure}
\includegraphics[width=3.0in,height=4.0in,clip]{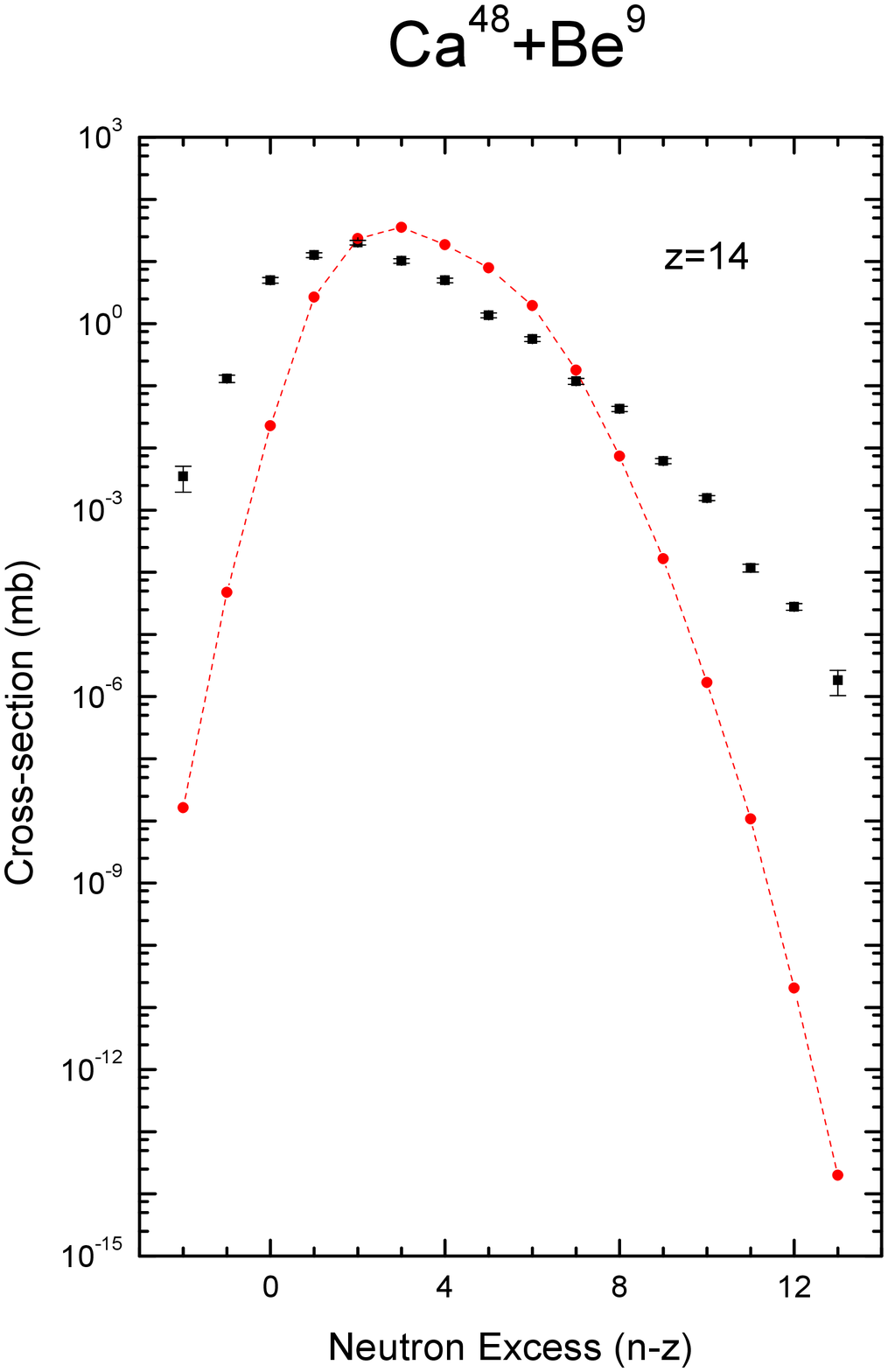}
\caption{Theoretical cross-section distribution (circles joined by
dashed line) of silicon isotopes for $^{48}$Ca on $^{9}$Be reaction
compared with experimental data (squares with error bars). The
theoretical calculation is done at a temperature T=4.25 MeV.}
\label{fig13}
\end{figure}

\end{document}